\documentstyle[12pt,equation,subeqn]{article}
\begin{document}
\newcommand{\tcr}{T_{cr}}
\newcommand{\df}{\delta \phi}
\newcommand{\drho}{\delta \rho}
\newcommand{\di}{\Delta I}
\newcommand{\uh}{\hat U}
\newcommand{\wt}{\tilde w}
\newcommand{\ct}{\tilde c}
\newcommand{\dur}{\frac{d \uh}{d \rho}}
\newcommand{\dus}{\frac{d \uh}{d \sigma}}
\newcommand{\durr}{\frac{d^{2} \uh}{d \rho^2}}
\newcommand{\duss}{\frac{d^{2} \uh}{d \sigma^2}}
\newcommand{\durs}{\frac{d^{2} \uh}{d \rho d \sigma}}
\newcommand{\sdur}{{d \uh}/{d \rho}}
\newcommand{\sdus}{{d \uh}/{d \sigma}}
\newcommand{\sdurr}{{d^{2} \uh}/{d \rho^2}}
\newcommand{\sduss}{{d^{2} \uh}/{d \sigma^2}}
\newcommand{\sdurs}{{d^{2} \uh}/{d \rho d \sigma}}
\newcommand{\dc}{\delta c}
\newcommand{\dchz}{\delta \chi_0}
\newcommand{\dchi}{\delta \chi_i}
\newcommand{\dkl}{\delta \kappa_{\Lambda}}
\newcommand{\lx}{\lambda}
\newcommand{\Lx}{\Lambda}
\newcommand{\Lt}{\tilde \Lambda}
\newcommand{\ex}{\epsilon}
\newcommand{\la}{\lambda_1}
\newcommand{\lb}{\lambda_2}
\newcommand{\gb}{\bar g}
\newcommand{\lab}{\bar{\lambda}_1}
\newcommand{\lbb}{\bar{\lambda}_2}
\newcommand{\lar}{{\lambda}_{1R}}
\newcommand{\lbr}{{\lambda}_{2R}}
\newcommand{\gr}{g_{R}}
\newcommand{\ma}{{m}^2_1}
\newcommand{\mb}{{m}^2_2}
\newcommand{\mar}{{m}^2_{1R}}
\newcommand{\mbr}{{m}^2_{2R}}
\newcommand{\larz}{{\lambda}_{1R0}}
\newcommand{\lbrz}{{\lambda}_{2R0}}
\newcommand{\grz}{g_{R0}}
\newcommand{\marz}{{m}^2_{1R0}}
\newcommand{\mbrz}{{m}^2_{2R0}}
\newcommand{\Ma}{{M}^2_1}
\newcommand{\Mb}{{M}^2_2}
\newcommand{\rhz}{\rho_0}
\newcommand{\sz}{\sigma_0}
\newcommand{\sn}{\sqrt{N}}
\newcommand{\be}{\begin{equation}}
\newcommand{\ee}{\end{equation}}
\newcommand{\een}{\end{subequations}}
\newcommand{\ben}{\begin{subequations}}
\newcommand{\beq}{\begin{eqalignno}}
\newcommand{\eeq}{\end{eqalignno}}
\renewcommand{\thefootnote}{\fnsymbol{footnote} }
\def\lesimm{\lower3pt\hbox{$\,\buildrel < \over \sim\,$}}

\pagestyle{empty}
OUTP--94--11P
\vspace{30mm}
\begin{center}
{{ \Large  \bf
High Temperature Phase Transitions \\
\vspace{2mm}
in Two-Scalar Theories \\
\vspace{4mm}
with Large $N$ Techniques 
}} \\
\vspace{15mm}
J.A. Adams and N. Tetradis \\
\vspace{5mm}
{\em Theoretical Physics, University of Oxford,\\
 1 Keble Road, Oxford OX1 3NP, U.K. } \\
\end{center}

\setlength{\baselineskip}{20pt}
\setlength{\textwidth}{13cm}
 
\vspace{4.cm}
\begin{abstract}
{
We consider a theory of 
a scalar one-component field
$\phi$ coupled to a scalar $N$-component field $\chi$. 
Using large $N$ techiques we calculate 
the effective potential in the leading order in $1/N$.
We show that this is equivalent to a resummation of
an infinite subclass of graphs 
in perturbation theory,
which involve fluctuations of the $\chi$ field only.
We study the temperature dependence of the expectation
value of the $\phi$ field and the resulting first and
second order phase transitions.
}
\end{abstract}
\clearpage
\setlength{\baselineskip}{15pt}
\setlength{\textwidth}{16cm}
\pagestyle{plain}
\setcounter{page}{1}

\newpage

\setcounter{equation}{0}
\renewcommand{\theequation}{{}\arabic{equation}}

The original argument of Kirzhnits and Linde 
\cite{orig} for the restoration of the electroweak 
symmetry at sufficiently high temperatures led to the 
development of the formalism \cite{doljac}-\cite{linde}
for the quantitative study 
of the behaviour of field theories at finite temperature. 
An area of application of this formalism is the 
early universe, where 
the necessary temperatures for a variety of phase 
transitions were presumably realized. 
An example which is relevant for this paper is
the transitions which can drive inflation \cite{inflation}.
The electroweak phase transition has attracted a great amount of
interest recently \cite{ew}, due to the possibility that it can 
create the necessary conditions for the
generation of the baryon asymmetry of the universe 
\cite{baryon}.
The incompatibility of the resulting predictions for the 
mass of the Higgs boson \cite{bound} with the experimental
bound has led to extensions of the standard model with 
additional scalar fields \cite{phicube}. 
However, the perturbative approach to high temperature phase
transitions, which was developed in refs. 
\cite{doljac}-\cite{linde} and used in subsequent studies,
has been shown 
to be insufficient for the reliable discussion of 
critical scalar fields \cite{transition}. 
For a reliable treatment of the phase transition for the 
$\phi^4$ scalar theory 
one has to resort 
either to the renormalization group
\cite{transition} or other non-perturbative methods,
such as large $N$ techniques \cite{avlargen}.
Multi-scalar theories can also be problematic in the context of 
the perturbative approach, and the reliability of the 
predictions for the high temperature transitions 
may be questionable. This is the motivation for this work, in
which we make use of the $1/N$ expansion in order to 
study the phase transitions in a model of a scalar one-component 
field $\phi$ coupled to a scalar $N$-component field
$\chi$. We wish to study the effect of the $\chi$
fluctuations at non-zero temperature on the expectation value
of $\phi$. We shall show that the leading result in the $1/N$ 
expansion is equivalent to the resummation of an infinite
subclass of perturbative contributions which involve $\chi$
fluctuations. We shall also study the effective potential
for $\phi$ at non-zero temperature and the possible 
phase transitions in dependence on the temperature.
\par
We consider a theory of two real scalar fields:
the one-component field $\phi$ and the $N$-component field
$\chi$. The classical action is 
invariant under a $Z_2 \times O(N)$ 
symmetry and has the form
\beq
S[\phi, \chi_a] =~& \int d^d x \bigl\{
\frac{1}{2} \partial_{\mu} \phi \partial^{\mu} \phi 
+\frac{1}{2} \partial_{\mu} \chi^{a} \partial^{\mu} \chi_{a} 
+\frac{1}{2} M_{1}^{2} \phi^{2} + \frac{1}{2} M_{2}^{2} \chi^{a} \chi_{a} 
\nonumber \\
&+\frac{1}{8} \lab \phi^{4} 
+\frac{1}{8} \lbb \left( \chi^{a} \chi_{a} \right)^2 
+ \frac{1}{4} \gb \phi^{2} \chi^{a} \chi_{a} \bigr\},
\label{twoone} \eeq
with $a = 0,...N-1$.
For the time being the dimensionality of the Euclidean
space-time $d$ is kept arbitrary. 
We wish to study the theory in the limit of large $N$. 
In order to implement the large $N$ approximation we 
first introduce an auxiliary field $C(x)$ 
and make use of the identity  
\be 
\exp \biggl\{ - \int d^d x \frac{1}{8} \lbb \left( \chi^a \chi_a \right)^2 
\biggr\} \sim
\int [{\cal D} c] 
\exp \biggl\{ \int d^d x \left( 
\frac{1}{2} N c^2 - 
\frac{1}{2} \sqrt{N \lbb} \chi^a \chi_a c 
\right) \biggr\}.
\label{twotwo} \ee
The partition function can now be written as
\be
Z(H,J_a) = \int [{\cal D} \phi] [{\cal D} \chi_a]  [{\cal D} c] 
\exp \biggl\{ - S'[\phi,\chi_a,c] + 
\int d^dx \left( H \phi + J_a \chi^a \right) \biggr\},
\label{twothree} \ee
where
\beq
S'[\phi,\chi_a,c] =~& \int d^d x \bigl\{
\frac{1}{2} \partial_{\mu} \phi \partial^{\mu} \phi 
+\frac{1}{2} \partial_{\mu} \chi^{a} \partial^{\mu} \chi_{a} 
+\frac{1}{2} M_{1}^{2} \phi^{2} + \frac{1}{2} 
M_{2}^{2} \chi^{a} \chi_{a} 
\nonumber \\
&+\frac{1}{8} \lab \phi^{4} 
+ \frac{1}{4} \gb \phi^{2} \chi^{a} \chi_{a} 
- \frac{1}{2} N c^2 
+ \frac{1}{2} \sqrt{N \lbb} \chi^a \chi_a c
\bigr\}.
\label{twofour} \eeq
The effective action is defined as the Legendre 
transform of the logarithm of the partition function and
can be evaluated using standard
methods \cite{colwein,jackiw}.  
Without loss of generality we consider expectation values
for the fields $\phi$ and $\chi_0$. For this reason we set 
$J_i=0$ for $i=1,...N-1$. 
A systematic expansion in powers of $N$ is obtained 
\cite{largen2,avlargen,jain}
by treating these expectation values as being ${\cal O} (\sn)$
\footnote{
This can be achieved through an appropriate rescaling of the
fields.}
and considering couplings ${\cal O} (1/N)$. In practice one 
uses shifted fields according to 
\beq
\phi &= \sn \Phi + \df 
~~~~~~~~~~~~~~~~~~~
c = C + \frac{\dc}{\sn}
\nonumber \\
\chi_0 &= \sn {\rm X} + \dchz 
~~~~~~~~~~~~~~~~~~~
\chi_i = \dchi~~~~~~~~~~i=1,...N-1  
\label{twofive} \eeq
and considers couplings which scale with $N$ as 
\be
\lab = \frac{\la}{N},~~~~~~~~~~ 
\lbb = \frac{\lb}{N},~~~~~~~~~~ 
\gb = \frac{g}{N}.
\label{twosix} \ee
The effective action $S_{eff}(\Phi,{\rm X},C)$ is calculated as a series 
in $1/N$ by evaluating the 
terms in the loop expansion which are proportional to
a given power of $1/N$. Finally the auxiliary field $C$ is eliminated
by its equation of motion
\be
\frac{\delta S_{eff}}{\delta C} = 0.
\label{twoseven} \ee
\par
The leading (of order $(1/N)^{-1}$) 
contribution to the effective potential is given by the
expression
\beq
\uh(\rho,\sigma) = \frac{U (\rho,\sigma)}{N} =~& \Ma \rho 
+ \frac{\la}{2}\rho^{2} +\Mb \sigma +  g \rho \sigma 
+ \sqrt{\lb} C \sigma - \frac{C^2}{2} \nonumber \\
&+ \frac{1}{2} \int_{\Lambda} \frac{d^d q}{2\pi^{d}} 
\ln \left(q^{2}+\Mb+ g\rho + \sqrt{\lb} C \right) 
\label{twoeight} \eeq
where we have defined 
\be
\rho = \frac{\Phi^{2}}{2},~~~~~~~~~~~~~\sigma = \frac{{\rm X}^{2}}{2},
\label{twonine} \ee
and an ultraviolet cutoff $\Lambda$ is implied for the momentum
integration. 
The auxilliary field is determined by its equation of motion
\be
\sqrt{\lb} \sigma - C + 
\frac{1}{2} \int_{\Lambda} \frac{d^{d}q}{2\pi^{d}} 
\frac{\sqrt{\lb}}{\left(q^{2}+ \Mb + g\rho + \sqrt{\lb} C \right)} = 0.
\label{twoten} \ee
The wave function renormalization does not receive any corrections 
at this order in $1/N$.
\par 
It is convenient to eliminate the auxiliary field from our
expressions.
For this reason we use the derivatives of $\uh(\rho, \sigma)$
with respect to $\rho$ and $\sigma$.
{}{}From eqs. (\ref{twoeight}), (\ref{twoten}) we obtain
\beq
\frac{d \uh}{d \rho} &= \Ma +\la \rho + g \sigma 
+ \frac{g}{2} I_1\left(\frac{d \uh}{d\sigma} \right) \label{twoeleven} \\
\frac{d \uh}{d \sigma} &= \Mb +\lb \sigma + g \rho 
+ \frac{\lb}{2} I_1\left(\frac{d \uh}{d\sigma} \right) \label{twotwelve} \\
\frac{d^{2} \uh}{d \sigma^{2}} &= \frac{\lb}{1 
+ \frac{\lb}{2} I_2\left(\dus \right)} 
\label{twothirteen} \\
\frac{d^{2} \uh}{d \rho d \sigma} &=  \frac{g}{1 
+ \frac{\lb}{2} I_2\left(\dus \right)} 
\label{twofourteen} \\
\frac{d^{2} \uh}{d \rho^{2}} &= \la 
+ \frac{g}{\lb}\left(\frac{d^{2} \uh}{d \rho d \sigma} -g \right) 
\label{twofifteen} 
\eeq 
where 
\be
I_n(w) = \int_{\Lx} \frac{d^d q}{(2 \pi)^d} \frac{1}{(q^2+w)^n}.
\label{twosixteen} \ee
Expressions for higher derivatives of the effective potential can
be obtained through differentiation of the above equations.
The following equations are satisfied by the derivatives of 
$\uh(\rho,\sigma)$
\beq
\Mb- \frac{\lb}{g} \Ma &= \dus - \frac{\lb}{g} \dur +
\left( \frac{\la \lb}{g} -g \right) \rho
\label{twoseventeen} \\
\frac{g}{\lb} &= {\durs} \big/ {\duss}
\label{twoeighteen} \\
\left( \frac{\la \lb}{g} -g \right) &=
\left( {\durr \duss} \big/ {\durs} - \durs \right).
\label{twonineteen} \eeq
It is interesting to interpret the expressions
(\ref{twoeleven})-(\ref{twofifteen}) in terms of perturbation
theory. The first two include the classical 
contributions to the mass terms, as well as the leading 
quantum corrections
coming from the summation of an infinite 
series of ``daisy'' and ``super-daisy'' graphs \cite{doljac}.
A typical example of ``super-daisy'' corrections
is presented in fig. 1a.
These leading corrections involve only the 
``Goldstone'' fields $\chi_i$, whose mass is equal to 
$d \uh / d \sigma$.
The next three expressions incorporate the leading 
corrections to
the quartic couplings. Eqs. (\ref{twothirteen})-(\ref{twofifteen})
can be expanded in a power series of the bare couplings
$\la$, $\lb$, $g$. The usual infinite series of 
``chain'' graphs is reproduced,
with each chain composed of one loop graphs involving 
two full $\chi_i$ field propagators. The form of these 
corrections is shown in fig. 1b, where the black circles 
denote $\chi_i$ field propagators incorporating the 
``super-daisy'' corrections. 
We can see, therefore, how the leading result in $1/N$ 
can be interpreted as a resummation of an infinite 
subclass of diagrams of perturbation theory. This 
subclass is dominant for large $N$, but it is not 
sufficient for the discussion of certain aspects of the
theory. For example,
since the fluctuations of the ``radial'' fields $\phi$, $\chi_0$
are not taken into account at this order in $1/N$, we 
do not expect to obtain a convex effective potential. 
Also, in the case of a second order phase 
transition, we do not expect to observe non-trivial behavior 
when the ``radial'' fields become critical.
Instead we expect mean field behavior associated with 
these fields. 
\par
{}{}From this point on we concentrate on the four-dimensional
theory. We first define the zero temperature theory in the
phase with spontaneous symmetry breaking. 
We are interested in studying the effect of the 
fluctuations of the $\chi$ fields on 
the expectation value of $\phi$.
For this purpose we choose a pattern of symmetry breaking 
which corresponds to a choice for the minimum of the 
potential ($\rhz,\sz$) such that $(\rhz \not= 0,\sz = 0)$. 
This choice preserves the $O(N)$ symmetry of the $\chi$ fields
while breaking the $Z_2$ associated with the $\phi$ field.
At the classical level, 
a sufficient condition for the potential to have 
a single minimum at $(\rhz \not= 0,\sz = 0)$
is $\Mb > g \rhz$. 
The integrals $I_n$ 
defined in eq. (\ref{twosixteen}) have been discussed extensively
in the literature \cite{doljac,avlargen} and we 
simply quote the results which are relevant to our
investigation. For $d=4$ 
$I_1(w)$ is given by
\be
I_1(w) = \frac{\Lambda^2}{16 \pi^2} 
+ \frac{w}{16 \pi^2} \ln \left( \frac{w}{\Lambda^2} \right),
\label{threeone} \ee
where we have assumed $\Lx^2 \gg w$. 
We recognize the 
quadratic and logarithmic divergences of the four-dimensional theory. 
The rest of $I_n$ can be obtained through differentiation
with respect to $w$. 
At any point $\rho$
the renormalized theory can be parametrized in terms 
of the masses of the $\phi$ and $\chi$ fields
\beq
\mar &= \dur(\rho,0) + 2 \rho \durr(\rho,0),
\label{threetwoa} \\
\mbr &= \dus(\rho,0) 
\label{threetwob} \eeq
respectively, 
and the quartic couplings
\beq
\lar &= \durr(\rho,0),
~~~~~~~~
\lbr = \duss(\rho,0),
~~~~~~~~
\gr = \durs(\rho,0). \label{threetwoe} \eeq
It is more convenient to use $d \uh / d \rho$ instead of $\mar$ for
the parametrization of the effective potential.  
The two are related
through eqs. (\ref{threetwoa}).
{}{}From eqs. (\ref{twoeleven})-(\ref{twofifteen}), (\ref{threeone}) 
we obtain 
\beq
\dur &= \Ma + \la \rho +  \frac{g}{32 \pi^2} \Lx^2
+ \frac{g}{32 \pi^2} \mbr \ln \left( \frac{\mbr}{\Lx^2} \right)
\label{threethree} \\
\mbr &= \Mb + g \rho +  \frac{\lb}{32 \pi^2} \Lx^2
+  \frac{\lb}{32 \pi^2} \mbr \ln \left( \frac{\mbr}{\Lx^2} \right)
\label{threefour} \\
\lbr &= \lb \left[ 1 
- \frac{\lb}{32 \pi^2} \ln \left( \frac{\mbr}{\Lt^2} \right)
\right]^{-1}
\label{threefive} \\
\gr &= g \left[ 1 
- \frac{\lb}{32 \pi^2} \ln \left( \frac{\mbr}{\Lt^2} \right)
\right]^{-1}
\label{threesix} \\
\lar &= \la + \frac{g}{\lb} (\gr -g),
\label{threeseven} \eeq
where $\Lt^2=\Lx^2/e$.
The above equations uniquely specify the renormalized parameters 
of the theory in terms of the bare ones. 
They indicate how the ultraviolet divergences 
can be absorbed in the definitions of the renormalized 
mass terms and couplings.
We also point out that the theory is well behaved in the
infrared, since the infrared logarithmic singularities
are cut off by the mass of the $\chi$ fields. 
The presence of the logarithms can give rise to the
Coleman-Weinberg mechanism for radiative symmetry 
breaking \cite{colwein}. If $\mbr$ is sufficiently small
the logarithm in eq. (\ref{threethree}) gives a negative 
contribution to 
$d \uh / d \rho$
which can lead to symmetry breaking.
We shall return to this point after removing the ultraviolet
divergences from our expressions. 
\par
We define 
the renormalized theory at zero temperature in the phase 
with spontaneous symmetry breaking 
in terms of the minimum of the effective potential $\rhz$
(where $d \uh / d \rho (\rhz,0)=0$) and the parameters
$\mbrz = d\uh /d\sigma(\rhz,0)$,
$\larz = d^2\uh /d\rho^2(\rhz,0)$, 
$\lbrz = d^2\uh /d\sigma^2(\rhz,0)$, 
$\grz = d^2\uh /d\rho\sigma(\rhz,0)$.
By making use of eqs. (\ref{threethree})-(\ref{threeseven}) we can 
relate the parameters at any other point $\rho \not= \rhz$
to those at the minimum. 
We find
\beq
\dur &= \larz(\rho - \rhz) + \frac{\grz}{32 \pi^2}
\biggl\{ - \mbr + \mbrz + \mbr \ln \left( 
\frac{\mbr}{\mbrz} \right) \biggr\} 
\label{threeeight} \\
\mbr &= \mbrz + \grz(\rho - \rhz) + \frac{\lbrz}{32 \pi^2}
\biggl\{ - \mbr + \mbrz + \mbr \ln \left( 
\frac{\mbr}{\mbrz} \right) \biggr\} 
\label{threenine} \\
\lbr &= \lbrz \left[ 1 
- \frac{\lbrz}{32 \pi^2} \ln \left( \frac{\mbr}{\mbrz} \right)
\right]^{-1}
\label{threeten} \\
\gr &= \grz \left[ 1 
- \frac{\lbrz}{32 \pi^2} \ln \left( \frac{\mbr}{\mbrz} \right)
\right]^{-1}
\label{threeeleven} \\
\lar &= \larz + \frac{\grz \gr}{32 \pi^2} 
\ln \left( \frac{\mbr}{\mbrz} \right).
\label{threetwelve} \eeq
The absence of the ultraviolet cutoff from the 
above relations
is a manifestation of the renormalizability of the theory.
In order to guarantee that the 
mass term $\mbr$ remains positive at any point $(\rho,0)$ 
we require
that $\mbrz \geq \grz \rhz$. 
We can now see how the Coleman-Weinberg mechanism for
symmetry breaking arises. For a choice of parameters
$\mbrz = \grz \rhz$, $\lbrz = 0$, $\larz = \grz^2 / 32 \pi^2$
the effective potential is given by
\be
\dur = \frac{\grz^2}{32 \pi^2} \rho \ln \left( \frac{\rho}{\rhz} 
\right).
\label{threethirteen} \ee
It is clear that in this region of parameter space 
the breaking of the symmetry is 
driven by the logarithm arising from the radiative corrections.
At non-zero temperature, we expect first order 
phase transitions to appear 
for theories with radiative symmetry breaking.
\par
In order to extend our discussion to the non-zero temperature problem 
we only need to recall that, in Euclidean formalism, 
non-zero temperature $T$ results 
in periodic boundary conditions in the time direction (for bosonic fields),
with periodicity $1/T$.
This leads to a discrete spectrum for the zero component of the momentum
$q_0$ 
and replaces the integration over $q_0$ by summation over the
discrete spectrum. 
As a result, 
eqs. (\ref{twoeleven})-(\ref{twofifteen}) remain valid (with $d=4$),
but eq. (\ref{twosixteen}) is replaced by
\be
I_n(w,T) = T \sum_m
\int_{\Lambda} \frac{d^3 \vec{q}}{(2 \pi)^3}
\frac{1}{(\vec{q}^2+4 \pi^2 m^2 T^2 + w)^n}.  
\label{fourthree} \ee
We can separate the non-zero temperature contribution to
the above expression by defining
\be
I_n(w,T) = I_n(w) + \di_n(w,T),
\label{fourfour} \ee
with $I_n(w)$ given by eq. (\ref{twosixteen}) (with $d=4$).
$\di_1(w,T)$ has been evaluated elsewhere  
\cite{doljac} and reads
\be
\di_1 \left( w,T \right) = 
\frac{T^2}{2\pi^2} \int_0^{\infty} dx 
\frac{x^2}{\sqrt{x^2 + \wt}} 
\frac{1}{\exp \left( \sqrt{x^2 + \wt} \right) -1},
\label{fourfive} \ee
where $\wt = w /T^2$. The rest of $\di_n(w,T)$
can be obtained through differentiation of the 
above expression. The first terms in a high temperature 
expansion (small $\wt$) of eq. (\ref{fourfive}) are 
\be
\di_1 \left( w,T \right) = 
\frac{T^2}{2 \pi^2} \biggl\{
\frac{\pi^2}{6} - \frac{\pi}{2} \sqrt{\wt} 
-\frac{1}{8} \wt \ln \left( \frac{\wt}{\ct^2} \right) ...
\biggr\}
\label{foursix} \ee
where $\ct = \exp \left( \frac{1}{2} +\ln(4 \pi)- \gamma \right)
\simeq 11.6$. 
\par
The effective potential is now temperature dependent.
Similarly to the zero temperature case, at any point $\rho$
we define the (temperature dependent)
masses of the $\phi$ and $\chi$ fields
\beq
\mar(T) &= \dur(\rho,0,T) + 2 \rho \durr(\rho,0,T),
\label{fourseven} \\
\mbr(T) &= \dus(\rho,0,T) 
\label{foureight} \eeq
respectively, 
and the quartic couplings
\beq
\lar(T) &= \durr(\rho,0,T),
~~~~~~~
\lbr(T) = \duss(\rho,0,T), 
~~~~~~~
\gr(T) = \durs(\rho,0,T). \label{foureleven} \eeq
Again for convenience we use $d \uh / d \rho (T)$ instead of $\mar(T)$.
Eqs. (\ref{threethree})-(\ref{threeseven}) are replaced by 
\beq
\dur(T) &= \Ma + \la \rho +  \frac{g}{32 \pi^2} \Lx^2
+ \frac{g}{2} \biggl\{
\frac{1}{16 \pi^2} \mbr(T) \ln \left( \frac{\mbr(T)}{\Lx^2} \right)
+ \di_1 \left( \mbr(T),T \right) \biggr\}
\nonumber \\
&~~
\label{fourtwelve} \\
\mbr(T) &= \Mb + g \rho +  \frac{\lb}{32 \pi^2} \Lx^2
+  \frac{\lb}{2} \biggl\{ 
\frac{1}{16 \pi^2} \mbr(T) \ln \left( \frac{\mbr(T)}{\Lx^2} \right)
+ \di_1 \left( \mbr(T),T \right) \biggr\}
\nonumber \\
&~~
\label{fourthirteen} \\
\lbr(T) &= \lb \left[ 1 
- \frac{\lb}{2} \biggl\{
\frac{1}{16 \pi^2} \ln \left( \frac{\mbr(T)}{\Lt^2} \right)
- \di_2 \left( \mbr(T),T \right) \biggr\}
\right]^{-1}
\label{fourfourteen} \\
\gr(T) &= g \left[ 1 
- \frac{\lb}{2} \biggl\{
\frac{1}{16 \pi^2} \ln \left( \frac{\mbr(T)}{\Lt^2} \right)
- \di_2 \left( \mbr(T),T \right) \biggr\}
\right]^{-1}
\label{fourfifteen} \\
\lar(T) &= \la + \frac{g}{\lb} (\gr(T) -g).
\label{foursixteen} \eeq
The final step is to relate the temperature dependent renormalized 
parameters at a point $\rho$ to the parameters at 
the minimum of the zero temperature effective potential $\rhz$. 
The calculation is straightforward and we find
\beq
\dur(T) = &\larz(\rho - \rhz) + \frac{\grz}{32 \pi^2}
\biggl\{ - \mbr(T) + \mbrz + \mbr(T) \ln \left( 
\frac{\mbr(T)}{\mbrz} \right) \biggr\} 
\nonumber \\
&+\frac{\grz}{2} \di_1 \left( \mbr(T),T \right)
\label{fourseventeen} \\
\mbr(T) = &\mbrz + \grz(\rho - \rhz) + \frac{\lbrz}{32 \pi^2}
\biggl\{ - \mbr(T) + \mbrz + \mbr(T) \ln \left( 
\frac{\mbr(T)}{\mbrz} \right) \biggr\} 
\nonumber \\
&+\frac{\lbrz}{2} \di_1 \left( \mbr(T),T \right)
\label{foureighteen} \\
\lbr(T) = &\lbrz \left[ 1 
- \frac{\lbrz}{32 \pi^2} \ln \left( \frac{\mbr(T)}{\mbrz} \right)
+\frac{\lbrz}{2} \di_2 \left( \mbr(T),T \right)
\right]^{-1}
\label{fournineteen} \\
\gr(T) = &\grz \left[ 1 
- \frac{\lbrz}{32 \pi^2} \ln \left( \frac{\mbr(T)}{\mbrz} \right)
+\frac{\lbrz}{2} \di_2 \left( \mbr(T),T \right)
\right]^{-1}
\label{fourtwenty} \\
\lar(T) = &\larz + \frac{\grz \gr(T)}{2}
\left[
\frac{1}{16 \pi^2} 
\ln \left( \frac{\mbr(T)}{\mbrz} \right)
- \di_2 \left( \mbr(T),T \right)
\right].
\label{fourtwentyone} \eeq
Eqs. (\ref{fourseventeen})-(\ref{fourtwentyone}) 
are the master equations for the study of the 
behavior of the theory at 
non-zero temperature. For a given 
zero temperature renormalized theory, as specified 
by the parameters $\rhz, \mbrz, \larz, \lbrz, \grz$, they 
encode all the information 
(in leading order in $1/N$)
related to phase change and
metastability at non-zero temperature.
\par
In order to identify the regions of parameter
space which lead to high temperature phase transitions of different 
order, it is instructive to study 
eqs. (\ref{fourseventeen})-(\ref{fourtwentyone}) analytically 
in some limiting cases. 
For $\lbrz = 0$, $\mbrz = \grz \rhz$, eq. (\ref{foureighteen}) 
gives for the renormalized $\chi$ field mass 
\be
\mbr(T) = \grz \rho.
\label{fiveone} \ee
As a result the radiative 
contributions of the $\chi$ fields to the 
effective potential involve a strong
$\rho$ dependence. 
We find 
\be
\dur(T) = \left( \larz - \frac{\grz^2}{32 \pi^2} \right)
(\rho - \rhz)
+ \frac{\grz^2}{32 \pi^2} \rho 
\ln \left( \frac{\rho}{\rhz} \right)
+\frac{\grz}{2} \di_1 \left( \grz \rho,T \right)
\label{fivetwo} \ee
Let us consider first the case $\larz = \grz^2/32 \pi^2$, which
corresponds to radiative symmetry breaking. 
The logarithmic 
term gives a negative contribution to 
$d \uh / d \rho (T)$.
Its effect is compensated by the 
positive high temperature contribution 
$\propto \di_1 \left( \grz \rho,T \right)$
which increases with temperature. 
There are two points 
of zero 
$d \uh / d \rho (T)$,
corresponding to the 
minimum of the potential at non-zero $\rhz(T)$ and the maximum of  
the barrier separating it from another minimum at zero.
For sufficiently high temperature the 
minimum at zero becomes the 
absolute minimum of the potential and the secondary one 
eventually disappears. The behaviour is characteristic 
of a first order phase transition \cite{linde,logs}. 
The size of the discontinuity of the order parameter is 
set by the logarithmic term and is 
\be
\drho = {\cal O}(\rhz). 
\label{fivethree} \ee
This classifies the transition as a strongly first order one.  
The high temperature expansion of eq. (\ref{foursix})
for $\di_1 \left( \grz \rho,T \right)$
is not adequate for the 
study of this case, since $\grz \drho/\tcr^2$ can be estimated 
to be larger than one. 
In the opposite limit 
$\larz \gg \grz^2/32 \pi^2$, the logarithmic term is a minor 
correction. 
Making use of the high 
temperature expansion of eq. (\ref{foursix}),
we rewrite eq. (\ref{fivetwo}) as 
\be
\dur(T) = 
\larz (\rho - \rhz) 
+\frac{\grz}{24} T^2  
- \frac{\grz^{3/2}}{8 \pi} T \sqrt{\rho}~... 
\label{fivefour} \ee
Taking into account the dominant contribution 
$\propto T^2$
would lead to a prediction for 
a second order phase transition. 
However, the second term in the high temperature expansion
gives again a negative contribution to 
$d \uh /d \rho (T)$, which 
results in a weakly first order transition \cite{phicube}.
The critical temperature is 
$\tcr^2/\rhz \simeq 24 \larz / \grz$ and the discontinuity
in $\rho$ is much smaller than $\rhz$
\be
\drho = {\cal O} \left( \frac{\grz^2}{32 \pi^2 \larz} \rhz \right),
\label{fivefive} \ee
justifying the use of the high temperature expansion. 
For $\lbrz = 0$, $\mbrz = \grz \rhz$, 
the phase transition remains first order for 
arbitrarily large values of $\larz$. 
Even though the picture seems consistent at this level of 
the $1/N$ expansion, serious complications appear 
at higher orders. 
{}From dimensional analysis it is 
expected that multi-loop corrections
to the effective potential which involve the $\phi$ field 
(and which have not been taken into account 
by the $1/N$ expansion so far) are proportional
to powers of $\larz T \big/ \sqrt{\mar(T)}$,
with the mass term defined
in eq. (\ref{fourseven}). These contributions are 
divergent near the critical temperature 
when the discontinuity in the order parameter 
and the mass term approach zero.
As a result they overwhelm and, therefore, invalidate 
the leading order result for weakly enough first order transitions.
An adequate treatment of the infrared problem must control the 
physics associated with the fluctuations of the ``radial''
field $\phi$. 
This was done in
ref. \cite{transition} through the
renormalization group approach for the $O(N)$-symmetric 
scalar theory. The $N=1$ theory was shown to 
have a second order transition, with an effective three-dimensional
critical behaviour. For our model it is reasonable to expect that
the fluctuations of the $\phi$ field can affect the order of a
transition, when this is predicted to be weakly first order
by the leading $1/N$ calculation.
As a result the first order character of the transition 
is not reliably established by the leading $1/N$ result
for large $\larz$, even for 
$\lbrz = 0$, $\mbrz = \grz \rhz$. Also, 
deviations from the above values for 
$\lbrz$ and $\mbrz$ lead to predictions of second order
phase transitions for sufficiently large $\larz$, even within
the leading $1/N$ calculation. This becomes apparent through 
the numerical study of the effective potential.
\par
For a given set of parameters of the zero temperature 
theory, we have solved eqs. (\ref{fourseventeen})
and (\ref{foureighteen}) for $d \uh /d \rho (T)$, 
using the full expression (\ref{fourfive}) for 
$\di_1 \left( \mbr(T), T \right)$.
The effective potential is obtained through numerical 
integration of the result. 
Two typical examples are presented in fig. 2, where 
$\uh (\rho,T)$ is plotted for various temperatures.
({}From this point on we omit the $\sigma$ dependence of the potential 
in our notation, since always $\sigma=0$.)
In the first example the zero temperature parameters are 
$\mbrz/\rhz = 1.55$, $\larz = 10^{-2}$, $\lbrz = 10^{-3}$,
$\grz = 1.5$ and a strongly first order transition 
is observed. 
In the second 
$\larz = 5 \times 10^{-2}$ while the other parameters have the 
same values. The strength of the 
first order transition is clearly diminished.
The effect of increasing $\larz$ on the strength of the first
order transition is more obvious in fig. 3, where we plot
the location of the minimum of the potential as a function of
temperature, for 
$\mbrz/\rhz = 1.55$, $\lbrz = 10^{-3}$, $\grz = 1.5$ 
and various values of $\larz$. 
The solid lines indicate the location of the minimum as long as
it corresponds to the true vacuum of the potential. The dashed ones 
indicate that a deeper minimum has appeared at zero, while we
are still following the location of a false vacuum.
At the point where the solid and
dashed lines meet the two minima have equal depth.
It is clear that the discontinuity in $\rho$ diminishes with 
increasing $\larz$. 
The line for $\larz=1.0$
corresponds to a second order phase transition.
This is due to the fact that the other zero temperature 
parameters deviate slightly from the values 
$\lbrz = 0$, $\mbrz = \grz \rhz$
which were considered in the
analytical discussion.
In figs. 4 and 5 we demonstrate the effect of the other parameters 
of the zero temperature theory on the strength of the first order 
transition. In fig. 4 it is shown that increasing 
$\mbrz$ reduces the strength of the first order transition, which 
eventually becomes second order. Similarly, fig. 5
shows that larger 
self-interactions for the $\chi$ fields result in 
more weakly first order transitions, which again turn second
order. We mention at this point, that the omission of the 
$\phi$ field fluctuations at this order in $1/N$ results in
mean field behaviour for the second order transitions of the
theory. If the temperature dependence of $\marz(T)$ in the symmetric
phase is
parametrized as $\marz(T) = d \uh /d \rho (0,T) 
\propto \left( T - \tcr \right)^{2 \nu}$,
the numerical solution of eqs. (\ref{fourseventeen}), 
(\ref{foureighteen})
gives $\nu = 0.5$ very close to the critical temperature.
\par
We conclude that a first order transition is
obtained when the temperature dependent mass 
$\mbr(T)$ of the 
$\chi$ fields has a strong dependence on $\rho$.
This is achieved for the zero temperature parameter range
$\mbrz \simeq \grz \rhz$, $\lbrz \simeq 0$.
Then the strength of the first order transition is maximum for 
$\larz \simeq \grz^2/32 \pi^2$ and diminishes for increasing 
$\larz$. Deviations from the above range of parameters 
reduce the dependence of $\mbr(T)$ on $\rho$
and consequently the strength of the transition, which
eventually becomes second order.
Only for $\mbrz = \grz \rhz$, $\lbrz = 0$ 
the phase transition is predicted to be first
order for arbitrarily large $\larz$. 
However, for the choice of parameters for which 
weakly first order 
transitions are predicted, 
the $\phi$ field fluctuations 
(which are not taken into account by the leading
$1/N$ calculation) become
important, as indicated by divergent contributions at 
higher orders of the $1/N$ expansion.
Studies based on the renormalizaton group approach 
\cite{transition,avlargen}
indicate that the incorporation of these fluctuations leads 
to second order transitions. As a result the nature of 
these transitions cannot be firmly established in the 
context of the 
$1/N$ expansion. 
The renormalizaton group becomes an indispensable tool 
for the resolution of these open questions.
Studies of two-scalar models with 
the use of the renormalization group 
are presented in refs. \cite{twoscal}.

\section*{Figures}

\renewcommand{\labelenumi}{Fig. \arabic{enumi}}
\begin{enumerate}
\item  
a) A typical example of the ``super-daisy'' corrections
summed by eqs. (\ref{twoeleven})-(\ref{twotwelve}).\\
b) An example of the ``chain'' graphs 
summed by eqs. (\ref{twothirteen})-(\ref{twofifteen}).
Black circles denote full $\chi_i$ field propagators. 
\item  
$\uh (\rho,T)$ at increasing temperatures
($T_5 > T_4 >T_3 > T_2 >T_1$),
for zero temperature parameters:\\
a) $\mbrz/\rhz = 1.55$, $\larz = 10^{-2}$, $\lbrz = 10^{-3}$,
$\grz = 1.5$. \\
b) $\mbrz/\rhz = 1.55$, $\larz = 5 \times 10^{-2}$, $\lbrz = 10^{-3}$,
$\grz = 1.5$. 
\item  
Position of the potential minimum against temperature 
for zero temperature parameters: 
$\mbrz/\rhz = 1.55$, 
$\lbrz =10^{-3}$, $\grz =1.5$, and 
$\larz = 0.01$, 0.05, 0.2, 0.5, 1.
Solid lines indicate positions of true vacua, 
dashed lines indicate positions of false vacua.
The circles indicate two minima of equal depth.
\item  
Position of the potential minimum against temperature 
for zero temperature parameters: 
$\larz = 10^{-2}$, $\lbrz =10^{-3}$, $\grz =1.5$,  
and $\mbrz/\rhz = 1.55$, 2.0, 2.75, 3.5, 4.5.
Solid lines indicate positions of true vacua, 
dashed lines indicate positions of false vacua.
The circles indicate two minima of equal depth.
\item  
Position of the potential minimum against temperature 
for zero temperature parameters: 
$\mbrz/\rhz = 1.55 \times 10^{-1}$, 
$\larz = 10^{-3}$, $\grz = 1.5 \times 10^{-1}$, and 
$\lbrz = 0.1$, 0.5, 1, 1.75, 2.5.
Solid lines indicate positions of true vacua, 
dashed lines indicate positions of false vacua.
The circles indicate two minima of equal depth.
\end{enumerate}

\begin{thebibliography}{99}


\bibitem{orig}
D.A. Kirzhnits and A.D. Linde, 
Phys. Lett. B {\bf 42}, 471 (1972).

\bibitem{doljac}
L. Dolan and R. Jackiw, Phys. Rev. D {\bf 9}, 3320 (1974).

\bibitem{weinberg}
S. Weinberg, Phys. Rev. D {\bf 9}, 3357 (1974).

\bibitem{linde}
D.A. Kirzhnits and A.D. Linde, 
JETP {\bf 40}, 628 (1974); Ann. Phys. {\bf 101}, 195 (1976).

\bibitem{inflation}
A.H. Guth, Phys. Rev. D {\bf 23}, 347 (1981);
A.D. Linde, Phys. Lett. B {\bf 108}, 389 (1982); 
A. Albrecht and P.J. Steinhardt, Phys. Rev. Lett. {\bf 48},
1220 (1982).

\bibitem{ew}
See the Proceedings of the NATO Advanced Research Workshop:
Electroweak physics and the early universe, Sintra, 1994 
(Plenum Press)
and references therein. 

\bibitem{baryon}
V.A. Kuzmin, V.A. Rubakov, and M.E. Shaposhnikov, Phys. Lett. 
B {\bf 155}, 36 (1985);
M.E. Shaposhnikov, Nucl. Phys. B {\bf 287}, 757 (1987); 
ibid {\bf  299}, 797 (1988).

\bibitem{bound} 
A.I. Bochkarev and M.E. Shaposhnikov, Mod. Phys. Lett. A 
{\bf 2}, 417 (1987).

\bibitem{phicube} 
A.I. Bochkarev, S.V. Kuzmin and M.E. Shaposhnikov, Phys. Lett. B 
{\bf 244}, 275 (1990); Phys. Rev. D {\bf 43}, 369 (1991);
N. Turok and J. Zadrozny, Nucl. Phys. B {\bf 369}, 729 (1992);
S. Myint, Phys. Lett. B {\bf 287}, 325 (1992); 
J.R. Espinosa, M. Quiros and F. Zwirner, Phys. Lett. B {\bf 307}, 106
(1993).

\bibitem{transition}
N. Tetradis and C. Wetterich, Nucl. Phys. B. {\bf 398}, 659 (1993);
preprint DESY-93-094, HD-THEP-93-28, to appear in Nucl. Phys. B;
Int. J. Mod. Phys. A {\bf 9}, 4029 (1994).

\bibitem{avlargen}
M. Reuter, N. Tetradis and C. Wetterich, 
Nucl. Phys. B {\bf 401}, 567 (1993).

\bibitem{colwein}
S. Coleman and E. Weinberg, Phys. Rev. D {\bf 7}, 1888 (1973).

\bibitem{jackiw}
R. Jackiw, Phys. Rev. D {\bf 9}, 1686 (1974).

\bibitem{largen2}
H.J. Schnitzer, Phys. Rev. D {\bf 10}, 1800 (1974);
S. Coleman, R. Jackiw and H.D. Politzer, Phys. Rev. D {\bf 10}, 2491 (1974);
R.G. Root, Phys. Rev. D {\bf 10} 3322 (1974).

\bibitem{jain}
V. Jain, Nucl. Phys. B {\bf 394}, 707 (1993);
H. Meyer-Ortmanns and A. Patkos, Phys. Lett. B {\bf 297}, 331 (1992).

\bibitem{logs}
J. Iliopoulos and N. Papanicolaou, Nucl. Phys. B {\bf 111}, 209 (1976);
A. Guth and E. Weinberg, Phys. Rev. Lett. {\bf 45}, 1131 (1980);
E. Witten, Nucl. Phys. B {\bf 177}, 477 (1981);
M. Sher, Phys. Rep. {\bf 179}, 273 (1989).

\bibitem{twoscal}
S. Bornholdt, N. Tetradis and C. Wetterich, 
preprints OUTP-94-13 P, HD-THEP-94-28, and
OUTP-94-14 P, HD-THEP-94-15; 
V. Jain and A. Papadopoulos, Phys. Lett. B {\bf 314}, 95 (1993);
M. Alford and J. March-Russell, Nucl. Phys. B {\bf 417}, 527 (1994).




\end{thebibliography}
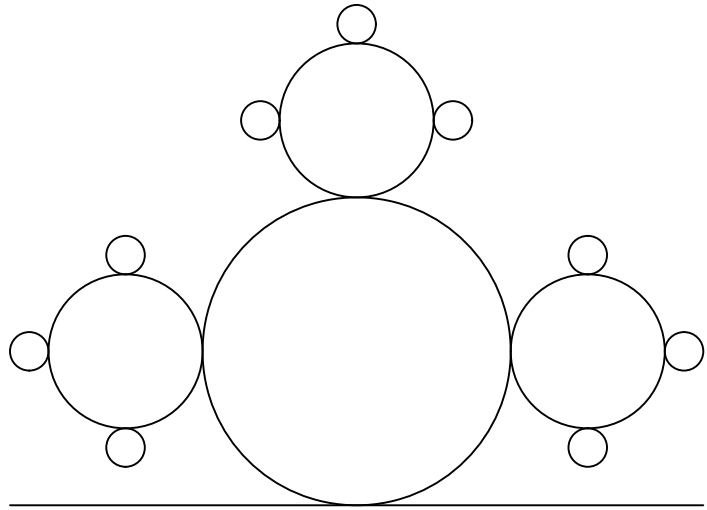
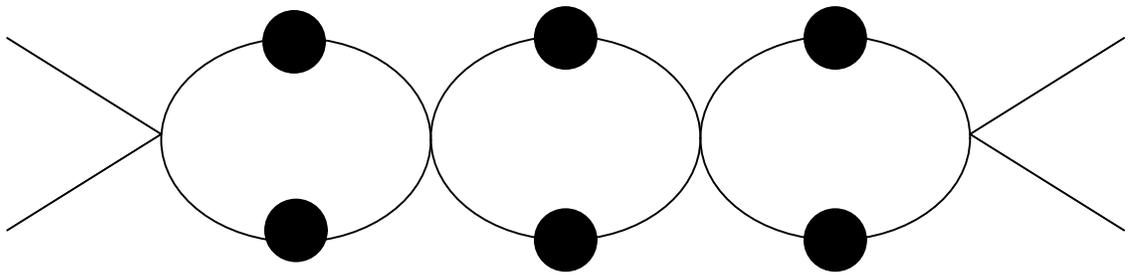
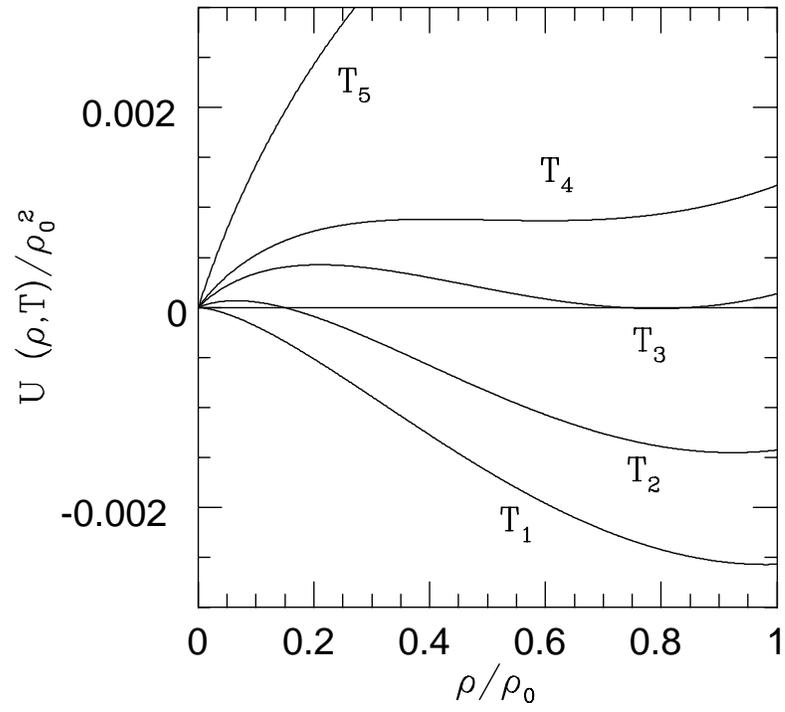
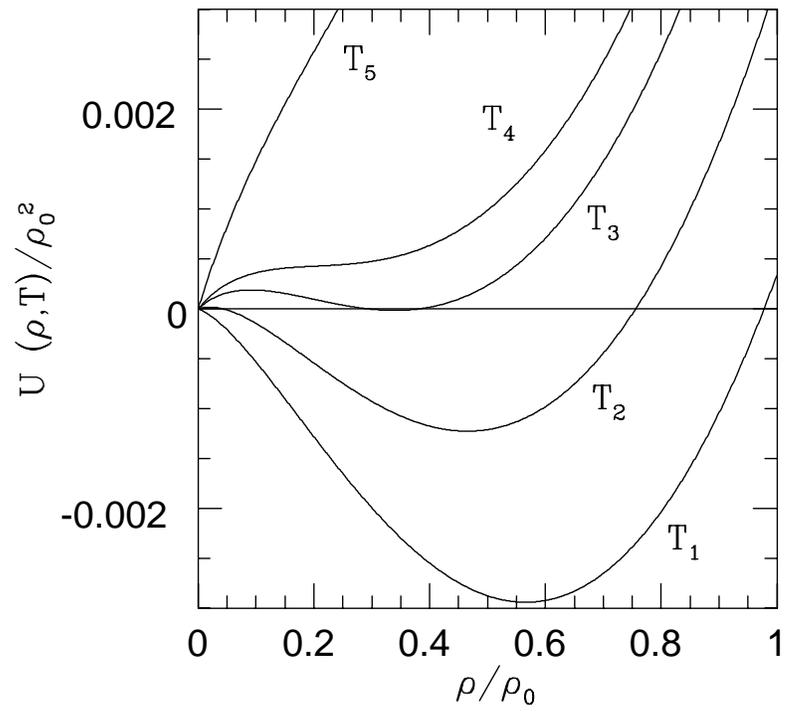
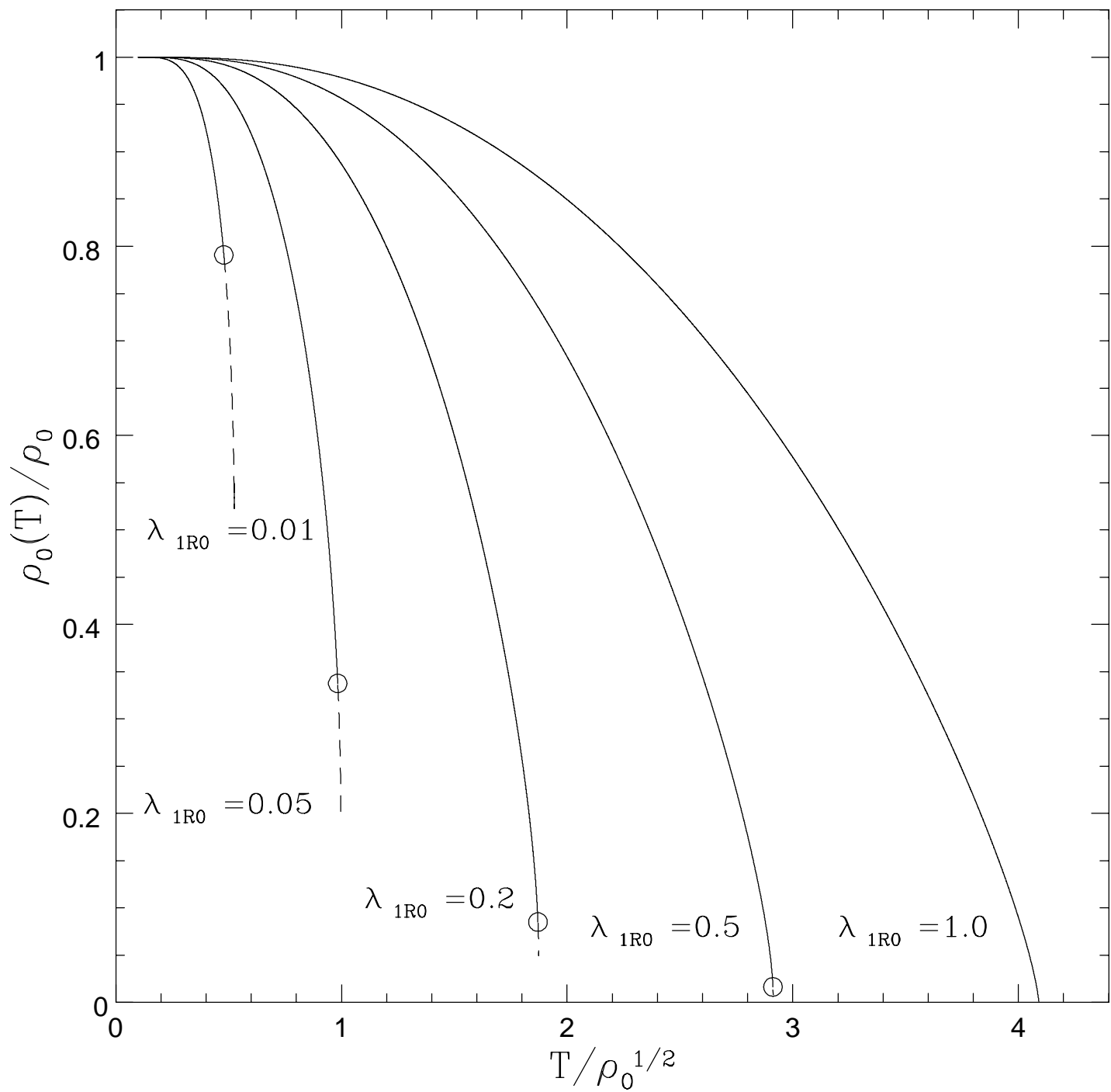
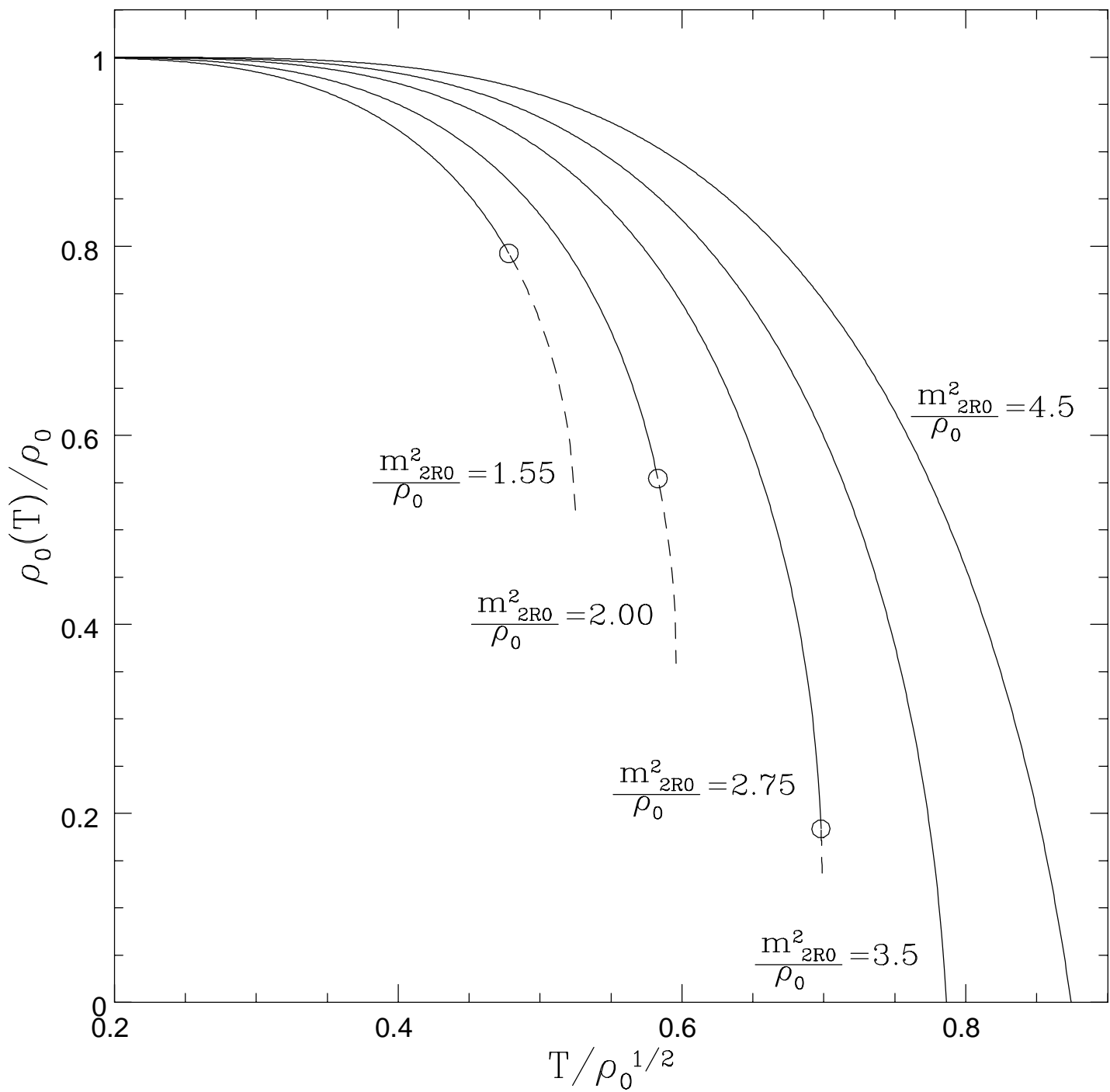
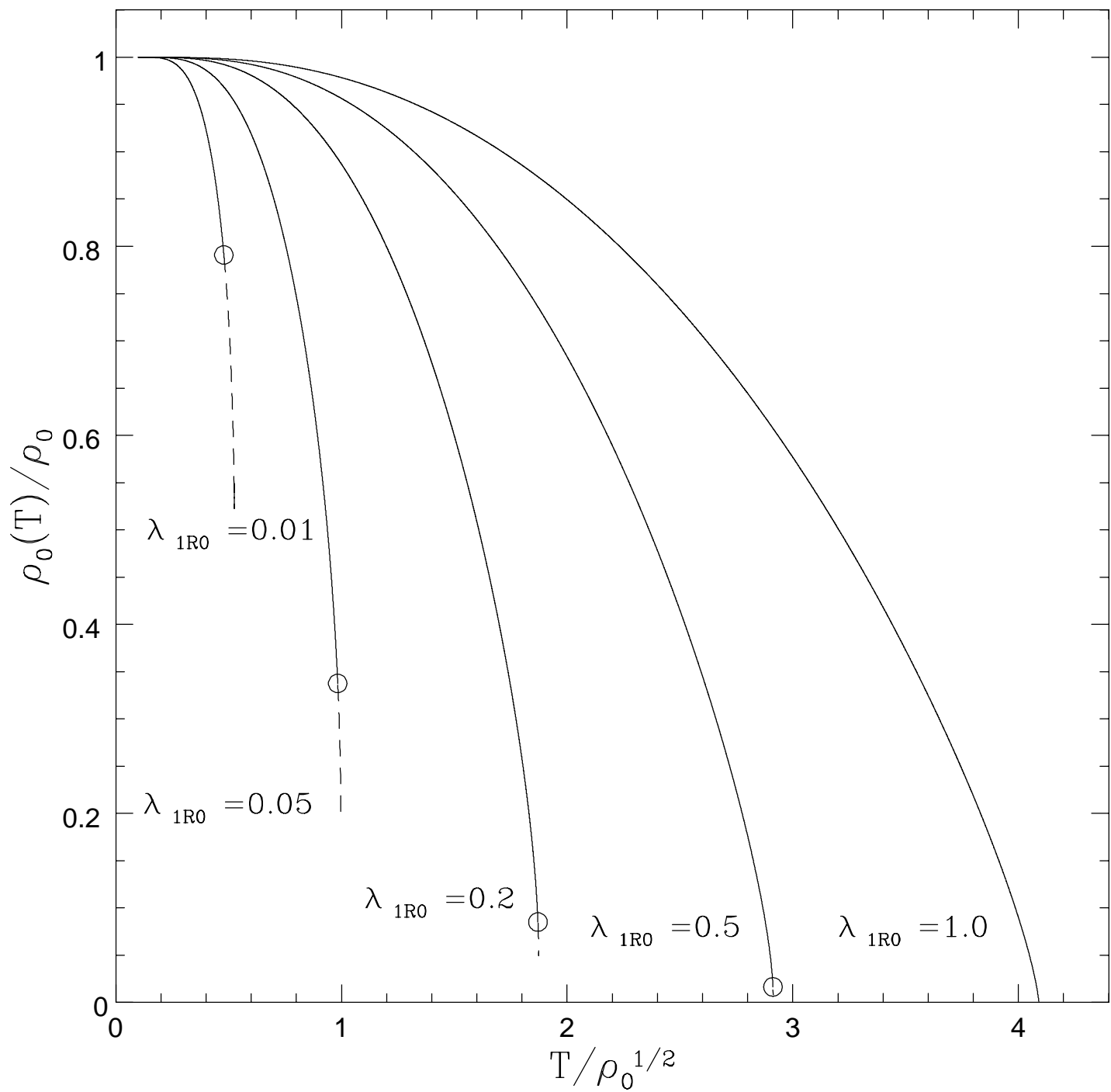
\end{document}